\documentclass[12pt]{article}
\usepackage[top=1.25in,bottom=1in,left=1in,right=1in]{geometry}
\usepackage[utf8]{inputenc}
\usepackage{amsmath}
\usepackage{natbib}
\usepackage[at]{easylist}
\usepackage{ascmac} 
\usepackage{amssymb}
\usepackage{amsthm}
\usepackage{bbm}
\usepackage{xcolor}
\usepackage{graphicx}
\usepackage{float}
\usepackage{enumerate}
\usepackage{natbib}
\usepackage{mathrsfs}
\usepackage{comment}
\usepackage{apptools}

\newcommand{\RN}[1]{  \textup{\uppercase\expandafter{\romannumeral#1}}}

\theoremstyle{remark}
\newtheorem{example}{Example}

\makeatother

\AtAppendix{
\numberwithin{equation}{section}
\numberwithin{definition}{section}
\numberwithin{theorem}{section}
\numberwithin{proposition}{section}
\numberwithin{lemma}{section}}

\DeclareMathOperator*{\argmin}{arg\,min}

\usepackage[most]{tcolorbox} 

\allowdisplaybreaks

\title{Using Machine Learning  to Generate, Clarify, and Improve Economic Models\thanks{I am grateful to David Romer and five anonymous referees for helpful feedback, and to Drew Fudenberg, Sendhil Mullainathan, Isaiah Andrews, Wayne Gao,  Jon Kleinberg, and Lihua Lei for many conversations that shaped my views on this topic.}}
\author{Annie Liang\thanks{Department of Economics, Northwestern University.}}
\date{\today}
\begin{document}

\maketitle

\section{Introduction}

Machine learning algorithms can now outperform classic economic models in predicting quantities ranging from  bargaining outcomes \citep*{CamererNaveSmith}  to choice under uncertainty \citep*{ellis2023predictivity} to an individual's future jobs and wages \citep*{vafa2024career}. Yet this predictive accuracy comes at a cost: most machine learning algorithms function as black boxes, offering little insight into \emph{why} outcomes occur.

This article asks whether machine learning can guide the development of new economic theories. Economic models serve an important purpose beyond prediction---they uncover the general mechanisms behind observed behaviors. A model that identifies the causal pathways of economic development is more valuable than one that merely predicts which countries will escape poverty, because it enables policymakers to encourage that development in countries where it might not have happened otherwise. Similarly, a model that predicts imperfectly across many domains can be more valuable than one that is highly accurate in a specific domain, since the former allows insights and data obtained from one setting to inform decisions and policy in another.

Applying machine learning algorithms off-the-shelf is unlikely to yield models with these properties. But a growing body of work shows that, when reconceived with the aims of an economic modeler in mind, machine learning methods can improve both prediction and understanding. These approaches range from adversarially training algorithms to expose the limits of existing models, to imposing economic theory as a constraint on algorithmic search. Recent advances in large language models  both complement these strategies and open entirely new research directions.

The plan of the paper is as follows. Section \ref{sec:Framework} introduces the setting of \emph{prediction problems}, in which the goal is to predict an unknown outcome $y \in Y$ from observables $x\in X$. A standard approach for learning a prediction rule $f: X \rightarrow Y$ is to specify a set of permissible mappings $\mathcal{F}$ and search for the mapping $f \in \mathcal{F}$ that best fits the available data. The key question is how to choose $\mathcal{F}$. Machine learning typically chooses large and expressive sets $\mathcal{F}$ that are computationally tractable to optimize over, while economics begins by restricting to a smaller $\mathcal{F}$ that embeds interpretable structure from the outset. As a result, machine learning algorithms approximate complex realities at the cost of interpretability, while economic modeling tends to preserve interpretability at the cost of failing to capture all the regularities in the data.

Sections \ref{sec:UnderstandModels}–\ref{sec:LLMs} form the core of the paper and explore different ways to integrate flexible machine learning approaches with more structured economic modeling. Section \ref{sec:UnderstandModels} argues that we can use machine learning to better understand our existing models---in particular, assessing how predictive, restrictive, and portable they are. This diagnostic step is important because we cannot improve our models without first understanding their limitations.

Section \ref{sec:ImproveModel} goes further and shows how machine learning can help improve existing models or generate new ones. The methods discussed here seek to uncover regularities our current models miss---either by studying how a more predictive black box algorithm beats the economic model, or by computationally probing the contours of the economic model and discovering where its predictions diverge from reality. The ideal output is a new model that retains interpretability while achieving greater predictive accuracy.

Section \ref{sec:Hybrid} generalizes the notion of an economic model to include hybrid approaches that are part economic theory and part black box. In particular, machine learning can be used to flexibly recover model primitives from high-dimensional data, while respecting the constraints that the economic theory imposes for how those model primitives are mapped into outcomes.

Section \ref{sec:LLMs} turns to large language models (LLMs) and the new opportunities they present. Early applications range from using LLMs  to stand in for human subjects in experiments, to using LLMs as a heuristic search tool over the space of ideas. I discuss how LLMs are qualitatively different from the supervised machine‐learning tools used in earlier papers, and point out some of the conceptual and methodological challenges they raise. 

Finally, Section \ref{sec:NotEcon} discusses related work in machine learning, in particular the literature on interpretable machine learning and on whether LLMs have learned ``world models.'' The motivation for this work is related to that of economic modeling, but there are also some important differences that I mention.

\bigskip

A number of excellent surveys examine the intersection of machine learning and economics, including \citet{mullainathan2017machine}, \citet{athey2019impact}, and \citet{athey2019machine} among others. These contributions have shaped the field’s understanding of how machine learning can improve prediction, inference, and policy evaluation in economics. This article differs in its focus---rather than treating ML primarily as a tool for empirical analysis, I consider how machine learning can be used to improve economic modeling.

\section{An Organizing Framework} \label{sec:Framework}

Section \ref{sec:PredictionProblem} introduces a  framework that encompasses common approaches in both machine learning and economic modeling. This framework is by no means exhaustive of questions in economics or of economic models,\footnote{
Some economic models are of a fundamentally different category from machine learning, and seek to distill relationships among concepts rather than making testable predictions about  economic outcomes \citep{Rubinstein2006}. This paper will instead exclusively consider testable economic models, which take fit to data seriously as a goal and are extended or modified when researchers identify disagreements between their predictions and actual behavior.} but it will focus our attention on a subset of problems and models where comparing machine learning and economic modeling is particularly productive. Section \ref{sec:OutofSample} reviews the practice of out-of-sample testing; readers familiar with this topic can skip ahead without loss.

\subsection{Prediction Problems} \label{sec:PredictionProblem}

An analyst seeks to predict an unknown outcome $y \in Y$ given an observable covariate vector $x \in X \subseteq \mathbb{R}^d$, where $x$ and $y$ are related by an unknown joint probability distribution $P$.  

\begin{example}[Risk Preferences] \label{ex:CE} An analyst wants to predict how individuals value uncertain prospects. Specifically, for a lottery $(z_1,p_1; \dots; z_n,p_n)$ paying $z_i$ with probability $p_i$, what certain payment $y$ would an individual consider equivalent?  The features $x$ describe the lottery's prizes and probabilities, and the outcome $y$ is the \emph{certainty equivalent}.
\end{example}

\begin{example}[Information Diffusion] The analyst would like to predict the extent of information diffusion across a social network, given initial seeding of the information with certain individuals. The features $x$ describe the underlying network and the individuals who are seeded with the information, and $y$ describes the fraction of the network that have heard about the product after $T$ periods.
\end{example}

To select a \emph{prediction rule} $f:X \rightarrow Y$, the analyst collects data $D=\{(x_i,y_i)\}_{i=1}^n$ and evaluates how well candidate functions $f$ fit that data as measured by a loss function $\ell: Y \times Y \rightarrow \mathbb{R}$. The rule that minimizes average loss on the data is 
\begin{equation} \label{eq:Optimum}
f^* \in \argmin_{f: X \rightarrow Y} \frac{1}{n} \sum_{(x,y) \in D} \ell(f(x),y)
\end{equation}
where  $\ell(f(x),y)$ quantifies the error to predicting $f(x)$ when the true outcome is $y$.\footnote{The minimum in  (\ref{eq:Optimum}) exists when (for example) the loss function is continuous and $\mathcal{X}$ and $\mathcal{Y}$ are compact. In (\ref{eq:ConstrainedOptimum}) below, also assume $\mathcal{F}$ is compact.}

The key choice is what set of candidate rules $f$ to consider in (\ref{eq:Optimum}), i.e., what function class to search over. Rather than optimizing over all possible maps $f: X \rightarrow Y$ as in (\ref{eq:Optimum}), the analyst typically restricts attention to a specific class of functions $\mathcal{F}$ and solves:
\begin{equation} \label{eq:ConstrainedOptimum}
f^* \in \argmin_{f \in \mathcal{F}} \frac{1}{n} \sum_{(x,y) \in D} \ell(f(x),y)
\end{equation}
This returns the prediction rule \emph{within} $\mathcal{F}$ that maximizes fit to the data. Thus if $\mathcal{F}$ is restricted to prediction rules with a particular structure---say, linear or monotone in $x_1$---the prediction rule $f^*$ will share that structure.

Traditional approaches in machine learning and economics differ in how the set $\mathcal{F}$ is chosen.

\paragraph{The machine learning choice of $\mathcal{F}$.}

In an idealized world with unlimited data and computing power, an analyst focused solely on accuracy would place no ex‐ante restrictions on $f$, giving the data complete freedom to speak. In reality, both data and compute are limited. But modern machine learning's success lies in identifying function classes $\mathcal{F}$ that are not only highly expressive---i.e., capable of approximating complex relationships---but also amenable to efficient optimization. Algorithms search within these large classes to find the function that best fits the data.

\begin{figure}[h]
 \begin{center}
     \includegraphics[scale=0.25]{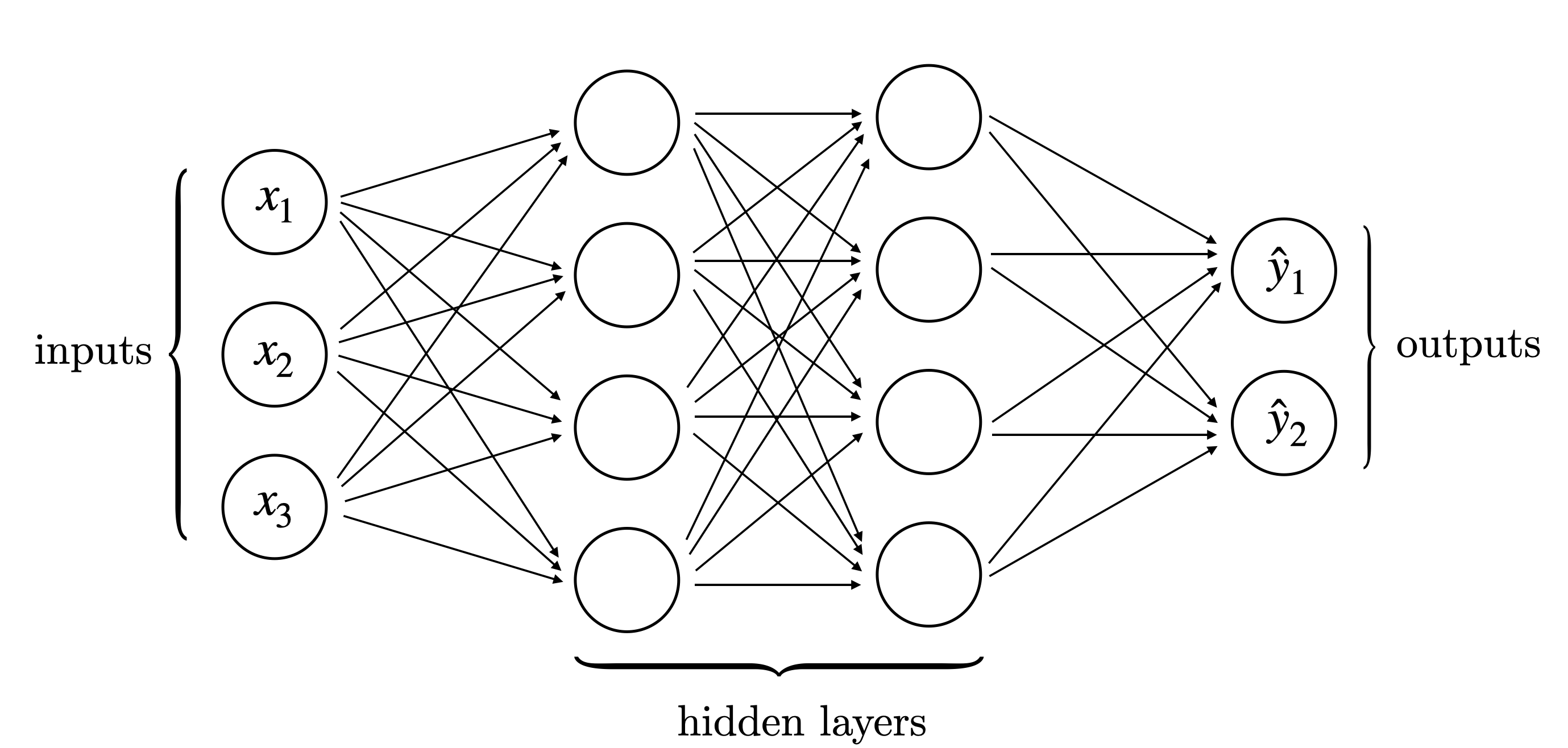}
 \end{center}
 \caption{\footnotesize{Example depiction of a neural network: the network processes input variables (leftmost layer) through a sequence of transformations, ultimately aggregating the final variables  into predictions (rightmost layer).}} \label{fig:NN}
 \end{figure}

Two widely-used machine learning algorithms are neural networks and random forests. The architecture of a neural network---depicted in Figure \ref{fig:NN}---consists of successive layers of interconnected ``neurons.'' Each neuron transforms the outputs from the preceding layer by applying a nonlinear activation function to a weighted sum of those outputs. The network successively propogates these transformed outputs through multiple layers, and in this way is able to accommodate increasingly complex functions of the original inputs. 

Neural networks are highly expressive: a network with even a single intermediate layer can approximate any continuous function on a compact domain, provided it has sufficiently many neurons \citep{HornikStinchcombeWhite1989}.  In this respect, the representational capacity of neural networks is similar to that of nonparametric estimators. Deep networks achieve similar representational power more efficiently by stacking multiple layers.

\begin{figure}[h]
\begin{center}
    \includegraphics[scale=0.32]{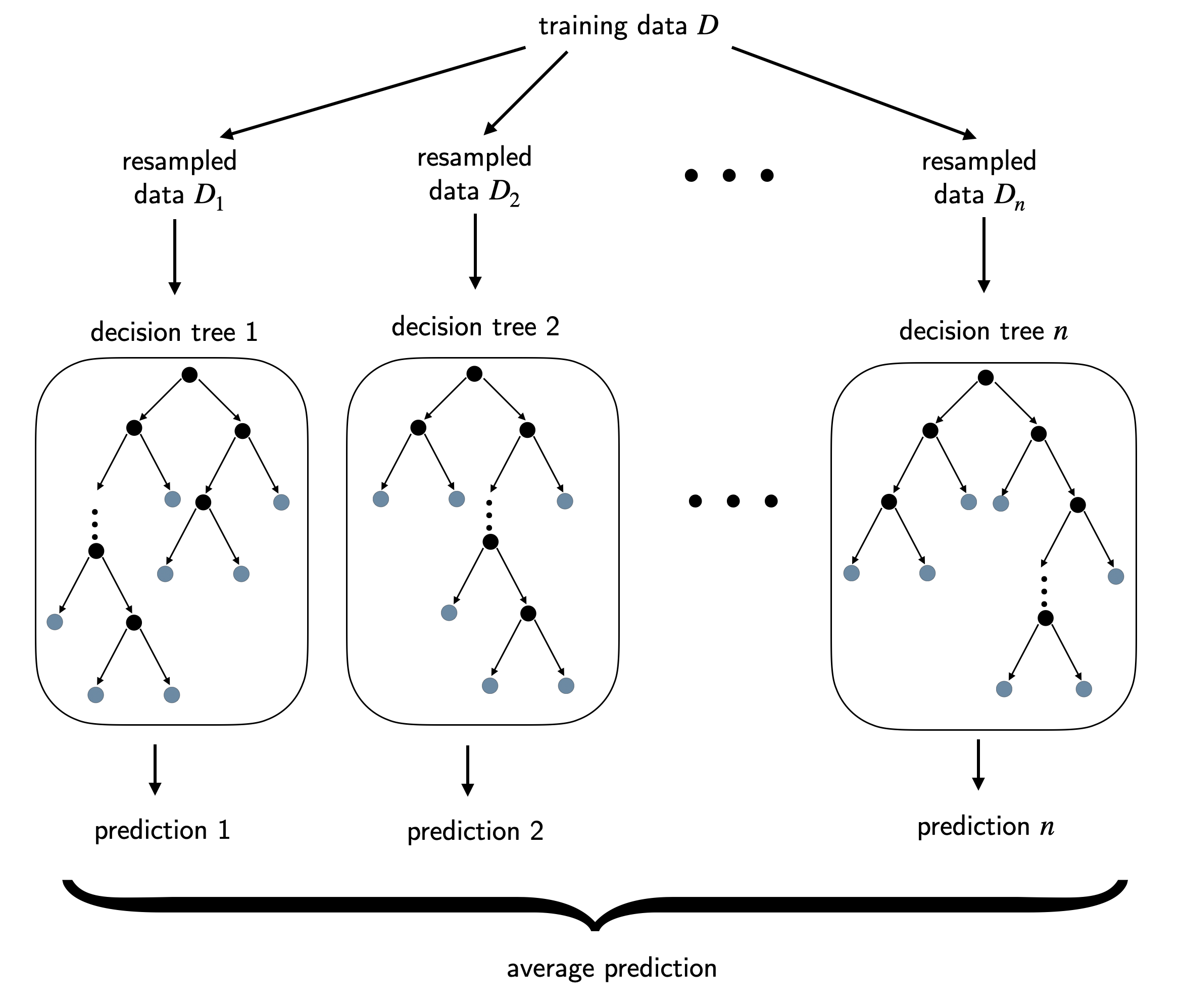}
    \caption{\footnotesize{Depiction of a random forest algorithm.}} \label{fig:RandomForest}
    \end{center}
\end{figure}

The random forest algorithm---depicted in Figure \ref{fig:RandomForest}---employs a different architecture but is also highly flexible. A random forest combines the predictions of an ensemble of decision trees, where each tree recursively partitions the feature space $X$, and assigns a constant prediction of $y$ to each resulting region---typically the average outcome for regression or the most frequent class for classification. The ensemble is constructed by repeatedly drawing bootstrap samples (with replacement) from the data, and training a separate decision tree on each sample.\footnote{To further reduce correlation among trees, it is common to consider a random subset of features at each split.} The random forest's prediction at  a new input is then the average of the predictions across trees.

Both algorithms can be applied to the problem in Example \ref{ex:CE} given data on lotteries and valuations. In this setting, a neural network would take the description of each lottery---i.e., the vector of prizes and their probabilities---and learn a nonlinear mapping into intermediate features, which are then transformed through successive layers to finally produce a prediction of the certainty‐equivalent.\footnote{Ideally, these intermediate features would be interpretable (e.g., capturing the lottery’s expected value or variance), but in practice they are typically not.} A random forest, by contrast, would iteratively partition the space of lotteries into regions defined by simple rules (e.g., ``the lottery does or does not have a negative prize'') and assign to each region a predicted certainty equivalent. The final prediction averages across many such trees. 

The use of these algorithms requires no prior knowledge about risk preferences or certainty equivalents, and yet with sufficient data they will produce highly accurate predictions. On the other hand, because the prediction rules learned by these algorithms are typically complex (multi-layered with many neurons per layer in the case of the neural net, and an ensemble of deep trees in the case of the random forest), they are difficult to interpret. While there are methods for deriving post-hoc explanations of specific predictions or aspects of a model’s structure (see Section \ref{sec:Interpretable}), the alternative approach in the following section begins by selecting a class of prediction rules that are globally interpretable by design.

\paragraph{The economic modeler's choice of $\mathcal{F}$.}

Sets $\mathcal{F}$ associated with economic models typically include only those prediction rules that possess some pre-specified  structure, thus guaranteeing that the selected prediction rule is interpretable.

Consider the setting of Example \ref{ex:CE}. One classic economic model of risk preferences posits that agents are expected utility maximizers with constant absolute risk-aversion (CARA), henceforth EU-CARA. The corresponding set $\mathcal{F}$ consists of all prediction rules $f: \mathbb{R}^n \times [0,1]^n \rightarrow \mathbb{R}$ of the form
\begin{equation} \label{eq:EU-CARA}f_\alpha(z_1,p_1,\dots,z_n,p_n) = \left(\sum_{0=1}^n p_i z_i^\alpha\right)^{\frac{1}{\alpha}}
\end{equation}
where the parameter $\alpha \in \mathbb{R}_+$ is a measure of the agent's risk aversion. This parameter is estimated from the data, and the selected prediction rule $f_{\alpha^*}$ is the one corresponding to the best-fit estimate $\alpha^*$. 

Relative to the machine learning algorithms discussed above, this class $\mathcal{F}$ imposes several restrictions on behavior. For example, EU-CARA rules out the possibility that individuals are expected utility maximizers but misperceive  probabilities. It also rules out the possibility that the sign of the prizes $z_i$ affects the size of the risk aversion coefficient $\alpha$. If actual behavior has those properties---so that EU-CARA is misspecified---then this model will not fit real data perfectly. On the other hand, the selected prediction rule is guaranteed to have a clear and economically interpretable meaning: the agent maximizes their expected utility, with utility a concave function of the dollar prize and $\alpha^*$ the risk aversion of the agent. 

\bigskip

These differences between EU-CARA and the random forest illustrate a general distinction between economic modeling and machine learning. The later sections ask whether algorithms such as the neural network or random forest can help us to better understand and build upon economic models such as EU-CARA. 

\subsection{Out-of-Sample Prediction} \label{sec:OutofSample}

The machine learning algorithms discussed above are (intentionally) very flexible. This allows the algorithms to search over a wide range of possible relationships between inputs and outputs, but it also raises the risk of \emph{overfitting}---producing a prediction rule that performs well on the data at hand but poorly on new, unseen data. To evaluate how well a model generalizes beyond the specific observations it was trained on, assessment of the performance of machine learning algorithms relies on \emph{out-of-sample prediction}.

In the simplest setting, out-of-sample prediction begins by splitting the available data into two disjoint sets: a \emph{training set} and a \emph{test set}. The training set is used to fit, or ``train,'' the model---that is, to select the specific prediction rule $f \in \mathcal{F}$ that solves (\ref{eq:ConstrainedOptimum}). The selected prediction rule is then evaluated on the test set: for each observation $(x_i, y_i)$ in the test set, the trained model produces a prediction $\hat{y}_i = f(x_i)$, and the prediction error $\ell(\hat{y}_i,y_i)$ is computed according to the chosen loss function. Because the test set was not used in fitting the model, this error can be used to estimate of the model's expected performance on new data drawn from the same distribution.\footnote{In practice, model training involves tuning \emph{hyperparameters}—settings such as the number of layers in a neural network or the maximum depth of a random forest—that affect model complexity and generalization. To select them without biasing test‐set error, a separate \emph{validation set} is used: models are trained on the training set, compared on the validation set, and the test set is reserved for final evaluation.}

A common variation on the single train-test sample split is \emph{$K$-fold cross-validation}. Here, the data is split into $K$ roughly equal parts (“folds”). The model is trained on $K-1$ folds and evaluated on the remaining fold, and this process is repeated $K$ times so that each fold serves as the test set exactly once. The resulting $K$ test errors are averaged to produce an overall estimate of out-of-sample performance. Cross-validation provides a more efficient use of limited data while still guarding against overfitting.

From the perspective of economic modeling, out-of-sample prediction plays a role analogous to testing a theory on a new dataset rather than on the one used to estimate it. For flexible machine learning algorithms such as neural networks and random forests, the discipline imposed by out-of-sample testing is essential: without it, the apparent accuracy of a model may be an artifact of its ability to memorize idiosyncrasies of the training data. The papers discussed in the following sections will use of out-of-sample prediction to evaluate the predictive performance of economic models and machine learning algorithms alike.\footnote{It is not strictly necessary to use out-of-sample testing to evaluate economic models with just a few parameters, and indeed the preceding literature has typically used in-sample estimators for the model's expected prediction error.}

\section{Using ML to Better Understand our Existing Models} \label{sec:UnderstandModels}

We expect economic models to be less predictive, more restrictive, and more portable than flexible machine learning algorithms. But is this actually the case? Assessing how well our models perform on these three dimensions is essential for understanding their strengths and limitations. This section introduces three computational metrics designed to evaluate these properties in economic models.

\subsection{Predictive Accuracy} \label{sec:Prediction} 

In most economic prediction problems, economic models fall well short of perfect accuracy. This can happen for two fundamentally different reasons. First, the model itself may fail to capture important relationships between the measured inputs and the outcome. For example, in the setting of Example 1, if individuals exhibit different risk preferences for losses and gains, or if they systematically misperceive probabilities, then EU–CARA will not predict certainty equivalents exactly.

A very different possibility is that the measured inputs themselves have inherently limited predictive power. For instance, if presenting the same lottery in different ways leads individuals to report different certainty equivalents, then \emph{no} model based solely on the lottery’s characteristics could predict the reported value perfectly all the time.

These two sources of error call for different approaches to model improvement. If a model performs substantially worse than the best achievable accuracy for its given covariates, then one might look for new models based on the same inputs. For example,  \citet{KahnemanTversky1979}'s Cumulative Prospect Theory relaxed Expected Utility to allow for misperceptions of probabilities and different treatment of gains and losses, and achieved better predictions given the same inputs.

By contrast, if the model is already close to the best possible performance for the existing feature set, then it is futile to attempt to improve prediction by specifying new functional forms based on the same inputs; better predictions require identifying and measuring new variables. In the context of Example 1, this might mean letting the model take as input the subject's demographic covariates or the framing of the lottery. 

One way to distinguish between these two error sources is to estimate the theoretical predictive limit given the measured covariates, i.e., the error of the prediction rule that solves the unconstrained problem (\ref{eq:Optimum}). This insight underlies \citet{PeysakhovichNaecker2017} and \citet{FKLM}.  Interestingly, for this goal, the flexibility of machine learning algorithms is not a drawback but rather an advantage. In fact, when the data contains sufficiently many $(x,y)$ observations  for each distinct value of $x$, then a simple ``lookup table'' algorithm that learns the best prediction of $y$ for each $x$ will approximate the best achievable error of any prediction rule $f: X\rightarrow Y$.  Such an algorithm thus provides a natural best‐case benchmark for evaluating how much room remains for improvement over an economic model.
 
\begin{figure}
\begin{center}
    \includegraphics[scale=0.28]{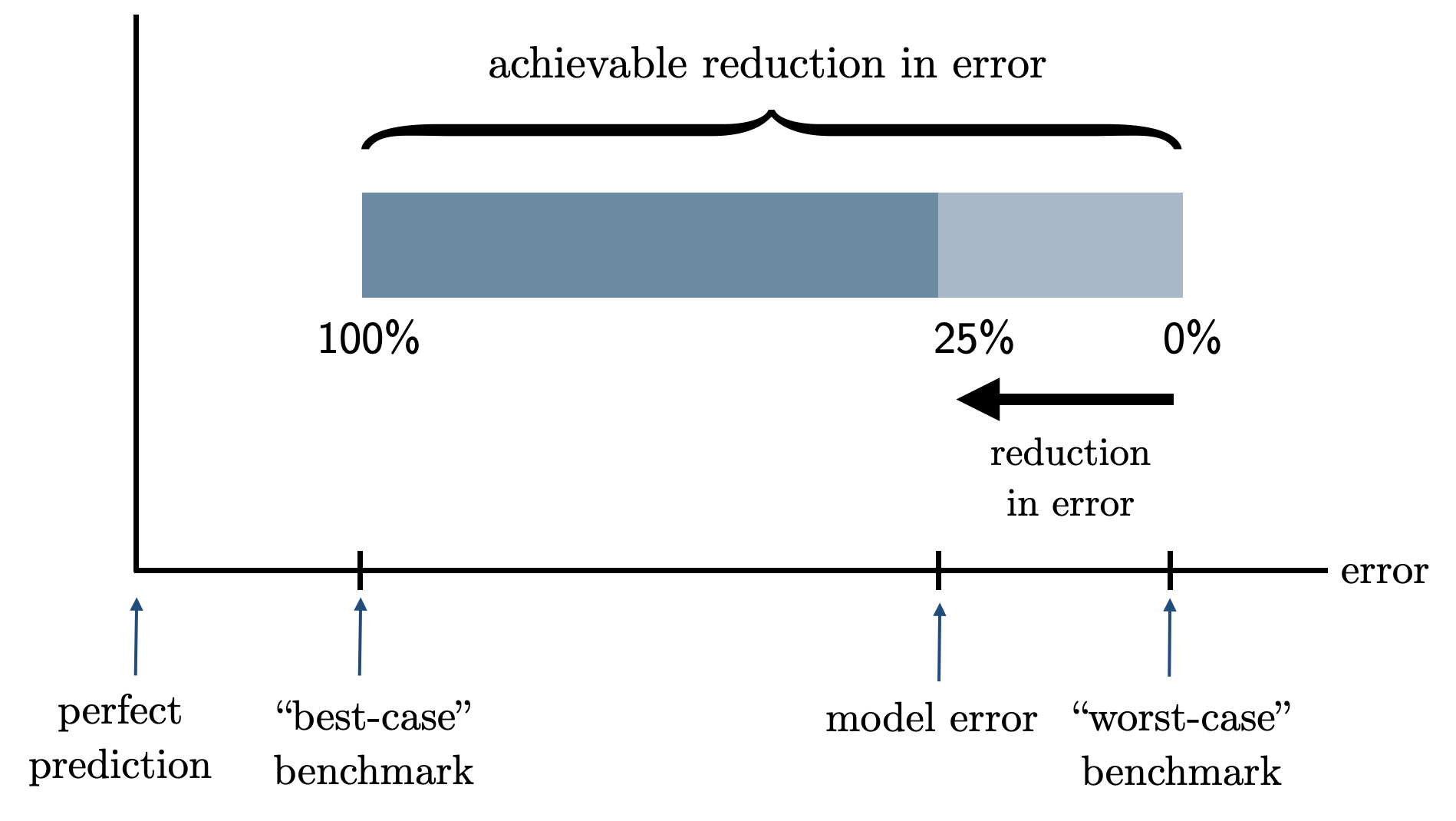}
    \caption{\footnotesize{A model's \emph{completeness} is the proportion of the gap between the worst-case and best-case prediction errors that the model is able to eliminate \citep{FKLM}. In the figure, the model's completeness is 0.25, indicating that the model reduces 25\% of achievable reduction in error from the worst-case to the best-case benchmark.}} \label{fig:Complete}
    \end{center}
\end{figure}

 \citet{FKLM} further propose using the error of a naive model (such as the expected value of the lottery in Example 1) as a worst-case benchmark, and then measuring where the economic model falls between these extremes. They define the \emph{completeness} of a model as the fraction of the gap between the worst-case and best-case benchmark that the model closes,
 \[\mbox{Completeness} = \frac{\mbox{Err}_{\mbox{\scriptsize{baseline}}} - \mbox{Err}_{\mbox{\scriptsize{model}}}}{\mbox{Err}_{\mbox{\scriptsize{baseline}}} - \mbox{Err}_{\mbox{\scriptsize{best}}}}\]
 as depicted in Figure \ref{fig:Complete}.

Although lookup table algorithms can estimate the best achievable error for a diverse range of laboratory data sets, they are not appropriate for most observational datasets.\footnote{In laboratory experiments, experimental economists control which problems are presented to subjects and can elicit many observations of $y$ for each $x$. This is rarely the case in naturally occurring data.} An alternative approach is to simply compare the performance of the economic model with that of another flexible machine learning algorithm, interpreting the latter as a proxy for the best achievable error in the problem. Several papers that do this find that economic models are nearly complete in their data. These settings include predicting certainty equivalents for two-outcome \citep*{FKLM} and three-outcome lotteries \citep*{PeysakhovichNaecker2017};  predicting choice between risky outcomes with two \citep*{ellis2023predictivity} and three \citep*{ellis2023demand} equiprobable states of the world; and predicting the average cooperation rate in play of the repeated Prisoner's Dilemma \citep*{fudenberg2024predicting}. These results do not rule out the possibility that machine learning algorithms would outperform the economic models given more data (although robustness checks such as varying the amount of data given to the machine learning algorithms can give us a sense of how valuable additional data would be). But they suggest that the economic models do not miss important regularities in these problems.

Other papers find that our existing models fall far short of the predictive accuracy of machine learning algorithms. The problems they consider include predicting the certainty equivalent for lotteries with unknown probabilities \citep*{PeysakhovichNaecker2017}; predicting the distribution of initial play of a normal-form game \citep*{HartfordWrightLeytonBrown}; predicting repeated play of a normal-form game \citep*{hirasawa2022using}; and predicting the distribution of choices between pairs of lotteries \citep*{Petersonetal}, among others. It is an interesting open question whether the problems and datasets for which economic models are complete can be systematically differentiated from the ones in which they are not.  

It is important to keep in mind that these results reflect performance in specific prediction tasks and specific problem instances (e.g., specific lotteries or games), and that a model's success in one domain does not necessarily translate to others. For example, a model that does well at predicting certainty equivalents for binary lotteries may perform less well with lotteries that have a larger support. Moreover, comparing economic models to machine learning algorithms in specific prediction problems presents an incomplete picture of performance, as economic models are intended to be useful across qualitatively different problems. We discuss this further in Section \ref{sec:Portability}.

There is thus need for future research that compares economic models and machine learning algorithms beyond narrow prediction tasks. This will require new comparative metrics that more comprehensively evaluate economic models against black box algorithms, such as average performance or variation in performance across a broad range of prediction problems. It will also require empirical work that more systematically explores the problem space,  revealing which types of problems our models handle well, and where they fall short. Section \ref{sec:Break} discusses how algorithms themselves might help guide this exploration.

\subsection{Restrictiveness} \label{sec:Restrictiveness}

The completeness metric from the preceding subsection quantifies how well a model predicts real data relative to the best achievable performance given the model's inputs. Yet completeness alone cannot tell us whether this success is because the model precisely captures the regularities inherent in this particular data, or because the model is very flexible and can (with sufficient data) flexibly adapt to capture most patterns. Indeed,  machine learning algorithms generally achieve high completeness for precisely the second reason.  To tell these cases apart, we need a complementary measure that captures how restrictive a model is.

\cite{selten1991properties} proposed a measure for the restrictiveness of an economic model based on the fraction of datasets it can perfectly explain. In the language of Section \ref{sec:Framework}, this is the fraction of all mappings $f: X \rightarrow Y$ that fall within the model's $\mathcal{F}$. A model that can explain any such mapping, given the right choice of parameter values, is completely unrestrictive.

\begin{figure}[h]
    \begin{center}
        \includegraphics[scale=0.3]{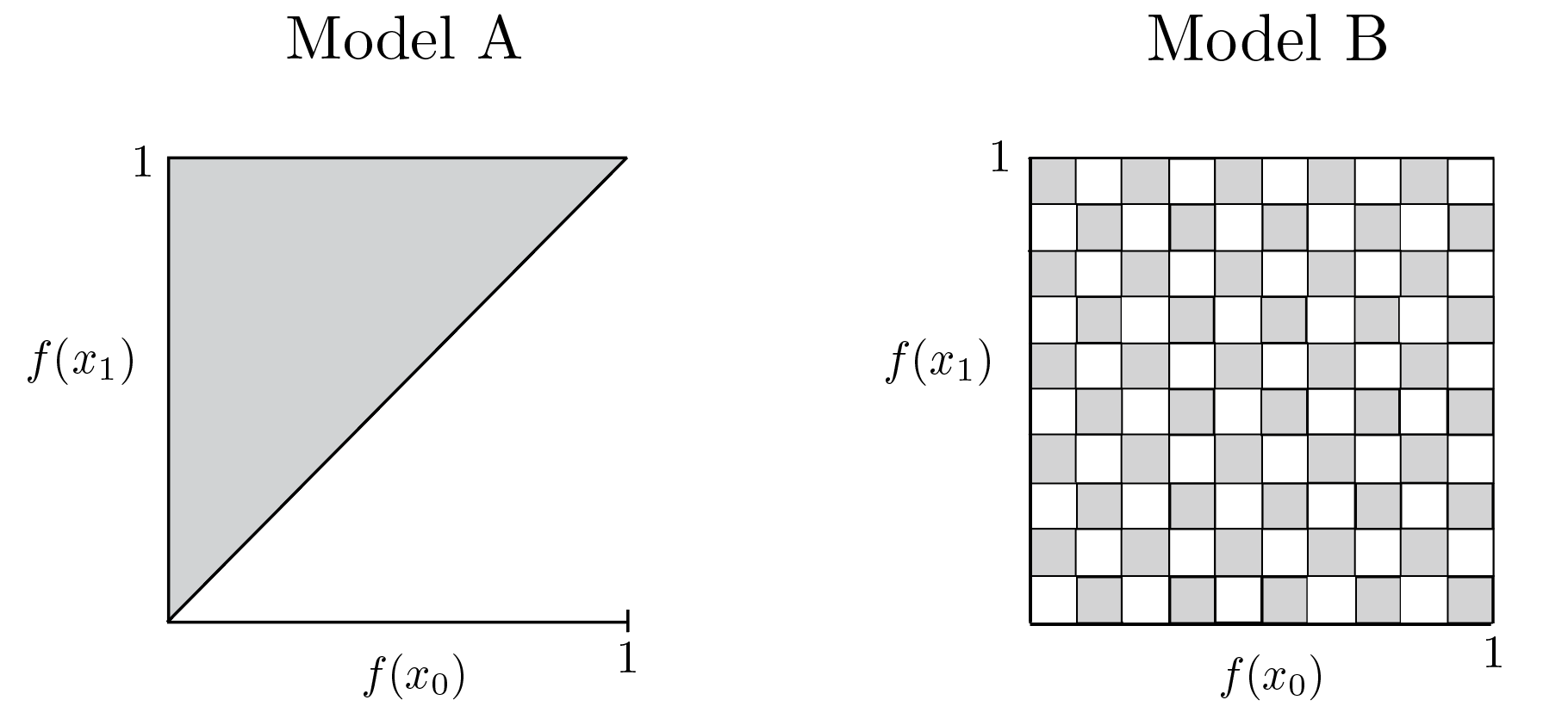}
        \caption{Models A and B are equally restrictive by the \citet{selten1991properties} measure.} \label{fig:Selten}
    \end{center}
\end{figure}

While intuitive, there are two challenges with applying this measure. The first is that it is generally difficult to determine whether a particular $f$ is consistent with a theory unless we already understand the theory's empirical content (e.g., by way of a representation theorem). Second, the \citet{selten1991properties} measure assigns identical scores to models that differ substantially in the kind of structure they impose. For a simple example, suppose there is a binary covariate $x\in\{x_0,x_1\}$ and an outcome $y \in [0,1]$. Figure \ref{fig:Selten} compares two models: Model A allows for all mappings $f$ that satisfy $f(x_1) > f(x_0)$, while Model B discretizes $[0,1]^2$ into a grid and allows for all mappings $f$ where the pair $(f(x_0),f(x_1))$ falls into the shaded set. In \citet{selten1991properties}’s metric, both models are equally restrictive (allowing for half of the possible mappings), yet Model A captures a clear structural regularity while Model B does not.

\citet{FudenbergGaoLiang} propose instead assessing restrictiveness as the average error a model makes when fitting randomly generated mappings $f$. They compute the model's best achievable fit to each of these mappings---formally, the smallest distance between the mapping and some $f \in \mathcal{F}$. These best-fit errors are averaged across many random draws and then normalized by the corresponding average error of a restrictive baseline model (such as predicting a constant). The resulting score lies between zero and one: a score of zero indicates that the model can perfectly approximate any mapping, while a score of one indicates that it approximates random mappings no better than the baseline.

Different from other classic measures of model restrictiveness such as the VC dimension, this measure does not require an analytical characterization of the model's implications and can thus be more easily applied to assess a wide range of models. In particular, \citet{FudenbergGaoLiang} apply their restrictiveness measure to evaluate a popular parametrization of Prospect Theory, which has been shown to have high completeness in predicting certainty equivalents for binary lotteries \citep{PeysakhovichNaecker2017,FKLM}. \citet{FudenbergGaoLiang} find that the model's restrictiveness on the domain of binary lotteries is low (a normalized score of 0.28). This means that the model's good fit to certainty equivalent data for binary lotteries should not be taken to mean that the model identifies the right restrictive structure, and could simply reflect flexibility. On the other hand, as the size of the lottery's support expands, the model imposes tighter restrictions on behavior. A good fit to data from these more complex lotteries would therefore provide stronger evidence that it captures the correct underlying structure.

Paired with completeness, restrictiveness locates models in a two‐dimensional space, and helps identify a Pareto frontier of models that are undominated in both respects. This joint view makes it possible to distinguish between predictive success that reflects meaningful theoretical structure and predictive success that may simply reflect flexibility. In particular, since adding a free parameter to a model necessarily improves completeness and reduces restrictiveness, researchers can compare the loss in restrictiveness to the improvement in predictive accuracy to assess whether the latter compensates for the former, as in \citet{Ba2022OverUnderreaction}.

 \begin{figure}[h]
 \begin{center}
     \includegraphics[scale=0.35]{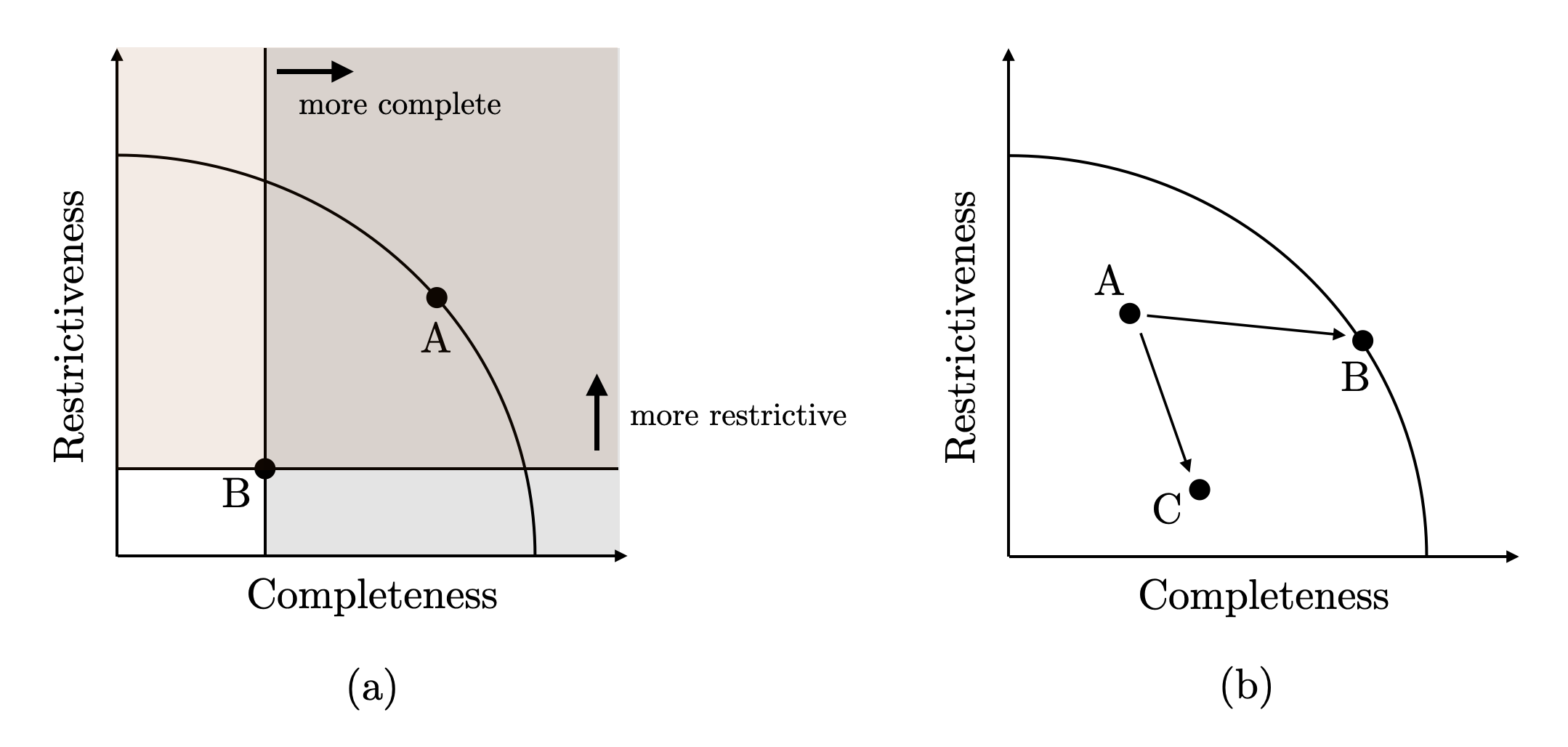}
 \end{center}
 \caption{\footnotesize{(a) Model A Pareto-dominates Model B by achieving both higher completeness and higher restrictiveness; that is, it simultaneously rules out more behaviors and also better explains real data; (b) Adding a free parameter to a model always involves a loss in restrictiveness and an improvement in completeness, but (all else equal) we prefer larger improvements in completeness and smaller reductions in restrictiveness, e.g., extending Model A to Model B rather than to Model C.}}

 \label{fig:Pareto}
 \end{figure}

Figure \ref{fig:Pareto} suggests an inherent tradeoff between restrictiveness and completeness. But in the extreme case where the true mapping $f$ for the prediction problem is already known, there is no real trade-off to consider: the singleton class $\{f\}$ is at once maximally restrictive and fully complete. A tension between the two objectives instead emerges in two common situations. First, the modeler may be uncertain about the data-generating process, and so must choose a larger hypothesis class 
$\mathcal{F}$ that has a good chance of containing the true $f$, even if this comes at the cost of greater flexibility. Second, the modeler may want the model to apply across different settings, where the relevant mapping changes from one environment to another, e.g., if all populations are risk averse but the risk aversion parameter varies. In both cases, expanding completeness means accommodating a wider range of plausible mappings. Developing a more precise theoretical foundation for this trade-off---one that articulates these practical goals and connects them back to the measures---remains an open direction for future work.

\subsection{Portability} \label{sec:Portability}

To this point, we have evaluated models within a single environment. Yet a central goal of economic modeling is to uncover structure that connects distinct economic domains.

A large literature addresses this challenge of \emph{out-of-domain generalization} (see \citet{zhou2021domain} for a survey). Unlike standard out-of-sample testing—where training and test sets are disjoint but drawn from the same distribution—out-of-domain generalization examines performance when training and test data come from different distributions. This is especially relevant for economics, where predictions are often needed in settings with no directly available data. For example, a development economist may need to forecast microfinance take-up in one Indian village using data from others, or an insurance economist may estimate willingness-to-pay for one set of insurance plans based on demand for a different set.

To systematically evaluate such cross-domain performance, \citet{andrews2024transfer} propose a general-purpose framework, which is applicable to both structural economic models and machine learning algorithms. They define a prediction method's \emph{transfer error} to be how well it performs when trained in one (randomly drawn) economic domain and used to make predictions in another (randomly drawn) domain. The approach requires the analyst to collect a meta-dataset consisting of samples across multiple domains. By pairing training and test samples from the observed domains, the analyst can estimate expected performance in an unobserved domain. The authors show how to construct simple confidence intervals for transfer performance under an assumption that the distributions governing the different domains are exchangeable, and further generalize these results when exchangeability is violated (but not by too much).

Applying this approach to models of risk preference, the authors find that black-box algorithms outperform economic models \emph{within domain} but have worse transfer performance across domains. Moreover, the economic model does not uniformly improve on the black box algorithm, but is instead advantaged precisely when extrapolation is required, i.e., when the training and test sets involve different lotteries. This finding echoes \citet{gechter2019evaluating}'s result that structural models produce better policy recommendations for conditional cash transfers in new contexts than black-box methods do. While \citet{andrews2024transfer} and \citet{gechter2019evaluating} provide evidence that economic models capture generalizable structure, additional empirical work is needed to establish how broadly this result applies.

Portability thus joins completeness and restrictiveness as a third measurable axis. Completeness captures the share of explainable variation a model accounts for within a domain; restrictiveness measures how much it rules out; and portability quantifies how well these gains carry across domains. 

\section{Using ML to Generate New Models and Hypotheses} \label{sec:ImproveModel}

When  machine learning algorithms outpredict our existing models, can we use these algorithms to uncover new regularities in behavior and to guide the extension of existing models? Several papers have made progress in exploring how algorithms can support the generation of new economic models and hypotheses. Below I discuss the successes and limitations of the ideas thus far.

\subsection{Adversarially Breaking Existing Models} \label{sec:Break}

To improve on a model, we should know where it goes wrong.

A powerful approach from computer science is to adversarially train an algorithm to ``break'' an existing model. For example, Generative Adversarial Networks (GANs) are a class of machine learning models that pit two neural networks---a generator and a discriminator---against each other in a game. The generator’s task is to create synthetic data (such as images) that mimic real data, while the discriminator’s role is to distinguish between real and generated data. As the generator improves at creating realistic data, the discriminator gets better at identifying fakes, driving both algorithms to improve over time.

 This adversarial approach is used in \citet{FudenbergLiang} to guide experimental design. They illustrate their method on \citet{StahlWilson94}'s Level-1 model of play in normal-form games. Although this model predicts existing experimental data very well, its good performance on these games does not imply that its performance will generalize to other games, since the games selected for lab experiments may share some special structure. To address this, \citet{FudenbergLiang} use machine learning to identify games on which the Level-1 model may fail. Using the existing experimental dataset, they train the algorithm that takes a game matrix as input and predicts the Level-1 model's predictive accuracy on that game. They then generate a large set of new games that the algorithm predicts will be challenging for the Level-1 model, and collect new data of play for these  ``algorithmically-generated'' games. They find that the Level-1 model indeed predicts behavior poorly in these games, and show that behavior in these games is not random but instead follows consistent patterns better explained by an alternative model. 

\citet{mullainathan2024automatic} similarly seek to identify examples on which a model fails. They frame such examples as \emph{anomalies}---instances, like the Allais paradox, that serve as an exemplar of a kind of behavior that could not be explained by the theory, and thus illuminate the predictive boundary of the theory.  They formalize the identification of anomalies as an adversarial game between a theory and a falsifier, where the falsifier proposes examples---i.e., collections of feature-outcome pairs---and the theory attempts to explain these examples by fitting its allowable functions to them. To solve for equilibrium of this game, they develop a gradient descent procedure that takes small steps in directions where the model's predictions change minimally, and yet actual behavior changes substantially. Applying this method to expected utility theory, the authors uncover known anomalies---such as the Allais paradox and the Common Ratio effect---as well as anomalies that have not been previously documented. They further verify that experimental subjects indeed violate expected utility theory on these algorithmically generated examples.

These papers demonstrate that machine learning can systematically expose the limitations of economic models, helping researchers refine theories by focusing on parts of the problem space where those models fall short. Designing environments that reveal a model's weaknesses can surface behavioral patterns that remain poorly understood. The challenge now is to extend these methods to moreover group and characterize the anomalies they uncover, enabling researchers to map not only where models break down, but also \emph{why}.

\subsection{Learning What the Machine Learning Models Get Right That the Economic Models Don't}

Another approach is to start with a more predictive machine learning algorithm and use human insight to interpret its edge over existing models.

\citet{FudenbergLiang} apply this strategy to predict play in normal-form games. They first show that a machine learning algorithm outperforms the leading economic models in out-of-sample prediction. They then focus on the subset of games where the machine learning algorithm is correct and the economic model's prediction is not. These games are interesting because the good performance of the machine learning algorithms suggests the presence of some underlying regularity, while the poor performance of the economic models suggests that this regularity is not included in the model. 

Examining these games reveals a common pattern: in each game, subjects responded to payoffs as if they were risk averse. The Level-1 model, which predicts that subjects choose the action that is a best response to uniform play by their opponent, does not allow for risk aversion, but can easily be extended to accommodate it. This new model matches the performance of the black box algorithms on the existing data, and continues to perform well on new data of play in other games, including in the later analysis of \citet{KULPMANN2022105554}.

\citet{hirasawa2022using} consider repeated play of a normal-form game with a unique (and complex) mixed strategy equilibrium, and use machine learning to reveal new behavioral regularities. They first compare the performance of \citet{CamererHo1999}'s Experience-Weighted Attraction model (henceforth EWA) against that of machine learning models, finding that EWA is only approximately 30\% complete. To understand which covariates play an important role in the machine learning algorithms, they consider feature importance measures (see Section \ref{sec:Interpretable}). Examining the important features reveals that players categorize the actions into two distinct classes. That is, rather than randomizing independently across periods, players negatively autocorrelate their actions within each class. \citet{hirasawa2022using} extend EWA to include variables indicating whether actions in the same class were recently chosen, and find that this modified behavioral model matches the performance of the most predictive machine learning algorithm.

Finally, \citet{LudwigMullainathan2024} address the challenge of generating new hypotheses, leveraging crowd wisdom to understand what the machine learning algorithm understands that the economic model does not. Specifically, they consider the problem of how judges make pre-trial detention decisions based on a defendant’s mugshot. To better interpret the algorithm's predictions, they employ a morphing algorithm, which creates synthetic mugshots that transition smoothly between faces with low and high predicted likelihoods of detention. This morphing process systematically alters only those facial features that the algorithm deems relevant to its predictions, while holding other attributes constant. By generating pairs of synthetic images that differ mainly in their predicted detention probabilities, the authors are able to isolate facial characteristics that influence the algorithm's decisions. Human subjects are then shown these morphed images and asked to describe the differences they perceive, thus assigning interpretable labels to the algorithm’s findings. This approach can  be iterated further by generating pairs of images that are not distinguished by the first feature, and yet have different predicted likelihoods of detention. Through this process the authors discover that ``well-groomed'' and ``heavy-faced'' defendants are more likely to be released.

These approaches all yield interpretable new directions in which existing models can be improved, and thus provide a proof of concept that machine learning can guide discovery of simple and portable models. On the other hand, these papers do not fully automate the process of that discovery---human expertise and intuition remain an essential part of how the algorithm's output is transformed into meaningful hypotheses and models that can be tested further. An open question is to what extent the final step can also be guided by algorithms. The recent development of large language models (LLMs) offers an opportunity to close that gap. Are LLMs capable of translating the complex patterns and relationships discovered by black-box algorithms into human-interpretable explanations and testable models? This is an empirical question: Since LLMs are trained on the existing literature, their ability to generate explanations may be (for now) confined by the boundaries of existing knowledge. Nevertheless, the ability of LLMs to communicate in ordinary language gives them a distinctive and potentially significant role in the further automation of scientific modeling. (See Section \ref{sec:LLMs} for further discussion.)

\section{Hybrid Econ-ML Models} \label{sec:Hybrid}

\subsection{Using ML to Flexibly Learn Model Primitives}

\begin{figure}[h]
    \begin{center}
        \includegraphics[scale=0.35]{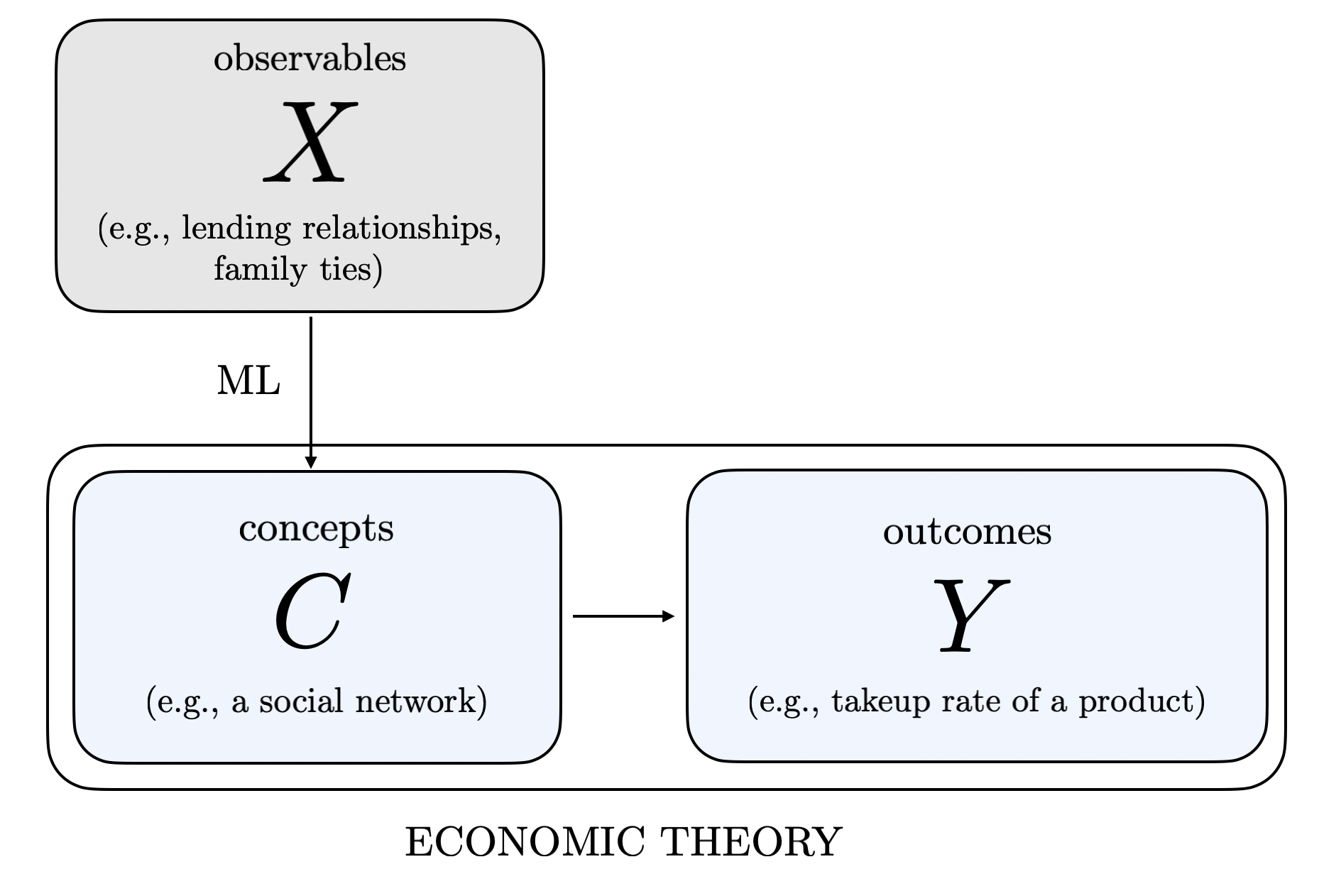}
        \caption{\small{Some economic theories can be viewed as a set of concepts $C$ and a set of maps $m:C \rightarrow Y$ relating concepts to outcomes. For example, a model of information diffusion might start with a social network as a primitive, and model how information is passed around the network, ultimately leading to a prediction of takeup rates. Such a model does not specify how we measure the social network, so to bring the model to data the researcher needs to further specify a map $\tau: X \rightarrow C$ taking observables into concepts.}} \label{fig:Diagram}
    \end{center}
\end{figure}

We have so far considered models that directly relate a set of observables $X$ to a set of outcomes $Y$. Many theories, however, posit intermediate concepts---model \emph{primitives}---that are not directly observable, such as the utility of a dollar prize or a network of relationships across individuals. These theories specify how the primitives relate to the outcome, but leave open how the primitives are derived from observables. This separation is often intentional, distinguishing the causal pathways we want to understand from the primitives we are willing to treat as given. See Figure \ref{fig:Diagram} for an illustration.

To apply such a model to data, researchers typically specify a functional form that links observables to the model primitives. But the predictive performance of the model, as assessed in this way, depends both on the quality of the model and the quality of the measurement of the model primitives. If the model predicts poorly, one possibility is mismeasurement of the primitives, rather than a failure of the model. The following papers apply machine learning to flexibly learn these primitives from the data.

\paragraph{Risky choice.}  \citet{Petersonetal} study individual choice between lotteries. Two canonical economic frameworks for this problem are Expected Utility Theory (EU) and Prospect Theory (PT). Under EU, an individual evaluates a gamble that pays $z_i$ with probability $p_i$ according to $\sum_i u(z_i)p_i$ for some utility function $u$. Prospect Theory generalizes this by transforming probabilities through a weighting function $\pi$, evaluating lotteries as $\sum_i u(z_i)\pi(p_i)$.

In empirical applications of these models, it is standard to impose parametric forms for $u$ and $\pi$, for example  $u(z_i)=z_i^\alpha$ (as in Section \ref{sec:Framework}) and $\pi(p)=\frac{\delta p^\gamma}{\delta p^\gamma + (1-p)^\gamma}$ (as proposed in \citet{goldstein1987expression}).  \citet{Petersonetal} instead allow a neural network to flexibly learn $u$ and $\pi$ from the data, referring to these specifications as neural EU and neural PT. Comparing neural EU with neural PT enables them to isolate the predictive gain from allowing probability weighting, without committing to any specific functional form for how probabilities are misperceived. They find that neural PT achieves slightly higher out-of-sample accuracy than neural EU, and also outperforms PT with the common parametric specifications of $\pi$ from the literature. A complementary direction for analysis would be to assess the restrictiveness of these models (Section \ref{sec:Restrictiveness}), thereby quantifying not only the gains in predictive accuracy but also the loss in model restrictiveness.

\paragraph{Salience.} Since \citet{schelling1960strategy}, the importance of salience in coordination games has been widely acknowledged, and yet rarely invoked in game theory due to the difficulty of measurement. \citet{Li2022Predictable} show that machine learning can quantify the salience of different actions, and that this quantity can be input directly  into game-theoretic models.

Consider hide and seek games in which two players---a hider and a seeker---are simultaneously asked to select a location from an image. The seeker wins if their locations match, and the hider wins otherwise. There is a unique Nash equilibrium in this game in which all locations are chosen equally often. But in empirical play, the seeker has an advantage relative to the equilibrium prediction, with matches happening more often than what uniform play would imply. 

To explain this, \citet{Li2022Predictable} propose a variation on a cognitive hierarchy model \citep{nagel1995unraveling,CamererHoChong04,crawford2007fatal}, which incorporates the idea of focal points. In this framework, level-$k$ players believe opponents are drawn from lower cognitive levels according to a Poisson distribution and  logit best-respond to these beliefs. The model is anchored by level-0 players, who are traditionally assumed to choose uniformly at random from the available actions. \citet{Li2022Predictable} instead suppose  that level-0 players choose locations with probability proportional to their visual salience, as quantified by the Salience Attention Model developed in computer vision \citep{vig2014large,cornia2018predicting}. This salience-augmented model better matches the empirical data.

\paragraph{Discrete choice.} In discrete choice models, the accuracy of predicted consumer behavior depends critically on how the products are characterized, but it is not obvious what the relevant characteristics are.

\citet{compiani2025demand} use pre-trained machine learning models to extract \emph{embeddings}---i.e.,  high-dimensional vector representations---from unstructured product image and text data.  They apply principal components analysis to these high-dimensional embeddings to recover the main dimensions of product differentiation, and incorporate these principal components in a mixed logit demand model \citep{mcfadden2000mixed,berry1995automobile}. This richer representation leads to more accurate recovery of substitution patterns across products, which in turn improves out-of-sample counterfactual predictions, such as predicting which products consumers would choose from a new choice set.

\citet{han2025copyright} adopt a related approach to study font markets and copyright policy. They represent each font as a point in a high-dimensional ``visual space'' derived from embeddings of font images, where distances in this space reflect perceived visual similarity. Firm competition is modeled as spatial differentiation within this space, where products that are closer together are closer substitutes. Within this framework, they model copyright protection as restricting entry to be at least a distance $d$ away from any existing font, and characterize how increasing $d$ affects firm entry decisions and consumer welfare. 

\paragraph{Parameter heterogeneity.} \citet{FarrellLiangMisra2021} considers settings in which an outcome $y$ depends on treatment variables $T$ through a model with parameters $\theta$. They assume that the structural relationship between 
$y$ and $T$ is common across the population, but that the parameters $\theta$ may vary with individual characteristics $X$. For example, the outcome may be linear in $T$ for all individuals, i.e., $y = \langle \theta, T\rangle$, while the coefficients $\theta$ differ across individuals. This generalization turns $\theta$ from a fixed vector into a potentially high-dimensional, nonlinear mapping $\theta(X)$, capturing heterogeneity not specified by the original theory.

To estimate these heterogeneous parameter functions, they design a \emph{structural deep neural network} in which the hidden layers feed into a ``parameter layer'' that outputs $\theta(X)$. This parameter layer is then passed through the fixed structural form---what they call the ``model layer.'' This architecture preserves the interpretability and counterfactual validity of the structural model, while exploiting deep learning’s flexibility to recover rich, nonlinear patterns in parameter heterogeneity.

\bigskip

There are several advantages to these ``machine-learning-augmented'' theories. First, they let us bring a broader range of theories to data by operationalizing concepts---such as salience or visual similarity---that were previously too vague or complex to quantify, thus expanding the set of theories we can formally test.

Second, they make existing theories more predictive by leveraging sources of complex and unstructured data that were previously ignored. Using this data, we can measure our model primitives more accurately, thereby improving the theory’s explanatory and predictive power.

Finally, these approaches help identify where further scientific progress is most needed. If optimizing the construction of primitives substantially improves predictive performance, this suggests that the bottleneck is measurement. And if, instead, increasingly sophisticated reconstruction of model primitives from data yields little predictive improvement, then it may be that the theory itself needs to be revised. 

\subsection{Using Economic Theory to Guide or Constrain ML}
Reversing the perspective of the previous section, we now consider how economic theory can inform the design of machine learning algorithms.

\begin{figure}[h]
    \begin{center}
        \includegraphics[scale=0.35]{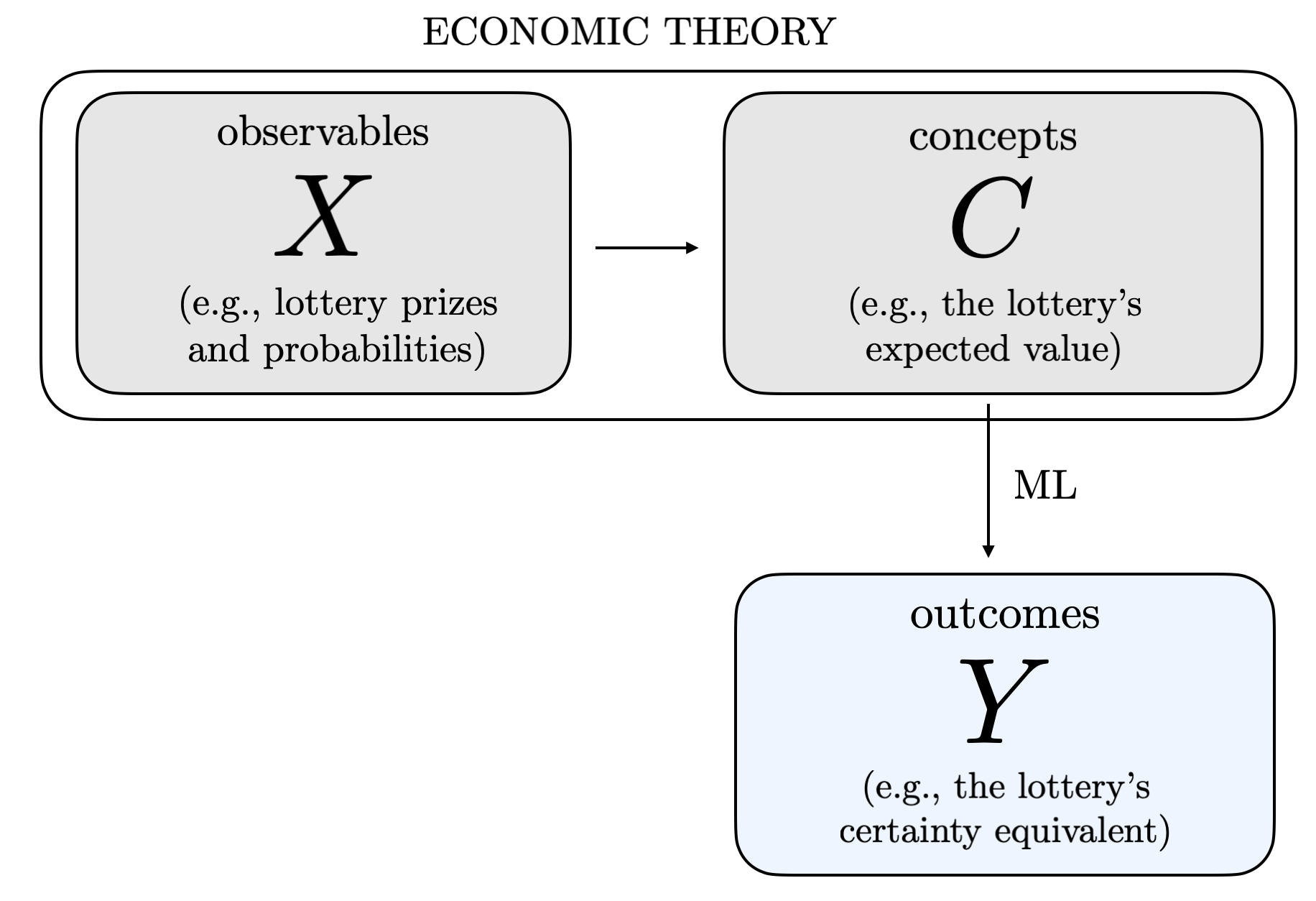}
        \caption{\small{Some theories identify relevant concepts, without specifying a precise form in which those concepts predict the outcome.}} \label{fig:DiagramInverse}
    \end{center}
\end{figure}

One way theory can contribute is by informing the selection and construction of features for the algorithm, a process often referred to as feature engineering. Consider the problem of predicting certainty equivalents for lotteries (Example \ref{ex:CE}). Given sufficient data, a flexible algorithm should (in principle) discover any relevant summary statistics of the lottery for this problem, such as its expected value or variance. In practice, when the search space is large and data are limited, performance can be improved by using theory to identify meaningful combinations of features and adding them explicitly to the feature set (see Figure \ref{fig:DiagramInverse}).


\citet{Plonskyetal2017} do precisely this. The problem they consider is predicting choice between two lotteries. They compare algorithms trained on three nested feature sets: The first set simply describes the two lotteries that are presented to the decision-maker, i.e., the prizes and their probabilities. The second feature set consists of the difference between the lotteries' expected values, the difference between their standard deviations, and the difference between their minimal and maximal outcomes.  The final feature set includes 13 additional features based on the preceding literature. These features include (among others) whether one lottery first-order stochastically dominates another, whether only losses are possible, and measures of regret.

Under ideal conditions---that is, if the model architecture is sufficiently flexible and it is trained on sufficient data---the information in the second and third feature sets is redundant, since those features can be reconstructed from the first feature set. Yet \citet{Plonskyetal2017} find that adding the basic features reduces mean squared error (MSE) by over 70\%, and that adding the psychological features yields a further 37\% reduction. Thus in this case, there is substantial predictive value to using theory to identify relevant features.\footnote{A similar approach is used in \cite{FudenbergLiang} and \citet{wright2014level} where domain knowledge is used to hand-craft features for predicting strategic play in matrix games.}

\bigskip

Moving beyond the use of theory to construct features (as depicted in Figure \ref{fig:DiagramInverse}), there are at least two additional ways in which economic theory can be used to guide algorithmic search.

\citet{HartfordWrightLeytonBrown} draw on economic theory to identify hard constraints on behavior, and encode these constraints as architectural invariances in a neural network. (It is well established that injecting domain knowledge of the input, particularly its topological structure, can improve representation learning \citep{bengio2013representation}. For example, \citet{clark2015training} exploit rotational invariance when training a Go-playing network.) Specifically, \citet{HartfordWrightLeytonBrown}  design a network to predict play in normal-form games, with the architecture hard-coding the symmetries implied by theory—for instance, relabeling either player’s actions leaves predictions unchanged except for a corresponding permutation of probability mass. They further implement iterative strategic reasoning through “action-response” layers, where each layer updates choices given beliefs from the layer below. The resulting model outperforms deep-learning baselines that rely on hand-crafted features.

By contrast, \citet{FesslerKasy} use economic theory not to impose hard constraints, but to discipline a prior over the space of models. They study the problem of learning a parameter vector $\beta$, where two extremes are to rely entirely on an unrestricted estimator or to adopt the predictions of a structural model. Rather than choosing between these, they propose a weighted average, with weights determined by how well the model’s predictions fit the data. Thus, instead of rigidly imposing the theory's predictions, this approach uses the predictions of economic models as an anchor for parameter estimation, shrinking estimators towards theoretical predictions when they are approximately correct, but reverting to an unrestricted model if the data diverge too much from the theory.

\section{Large language models} \label{sec:LLMs}

Large language models (LLMs) represent a qualitative break from the machine learning tools discussed so far. First, in contrast to  traditional supervised learning algorithms that are designed to map a fixed set of inputs $x$ to a specific output $y$, LLMs are general-purpose text generators---they can accept virtually any text as input and produce virtually any text in response. 

A second important difference is that LLMs operate in natural language, and thus can participate directly in the conceptual dialogue that drives theory construction. LLMs can process verbal arguments about causality, translate between mathematical formalism and economic intuition, and link abstract models to real-world phenomena. Thus, they have the potential to move beyond computing optimal solutions or estimating parameters, toward proposing mechanisms, identifying assumptions, and articulating the economic logic that connects primitives to predictions.

How best to use LLMs to augment economic modeling is an emerging area of research, with many of the most intriguing possibilities still being explored. Nevertheless, some unifying ideas are beginning to take shape. Section \ref{sec:LLMsasSubjects} presents LLMs as synthetic behavioral subjects capable of simulating human decision‐making, and Section \ref{sec:Heuristic} examines their potential to automate certain heuristic and creative aspects of research. Together, these perspectives suggest promising ways to accelerate and improve economics research, while also introducing distinct methodological and conceptual challenges.

\subsection{LLMs as Behavioral Subjects} \label{sec:LLMsasSubjects}
\citet{Horton2023HomoSilicus} argues that  LLMs can be viewed as ``computational models of humans'' that approximate human reasoning and bias sufficiently well to stand in as synthetic subjects in economic experiments. 

This is appealing for several reasons. First, there is an obvious cost advantage relative to recruiting and paying  human subjects. Second, 
the researcher can run a broader range of experiments, including (for now) ones that would be infeasible or unethical to conduct with human participants.\footnote{This may change as research and thinking progress on the ethics of interacting with AI.} For example, \citet{pmlr-v202-aher23a} use LLMs to simulate an obedience study based on the Milgram shock experiment. Third, LLMs potentially offer unprecedented control over the subject pool: Researchers can systematically vary traits such as age, gender, cultural background, political affiliation, cognitive style, or even risk tolerance. While further research is needed to assess how successfully identity can be elicited and how precisely users can control this process, this approach has the potential to generate precisely tailored subject profiles, enabling researchers to query virtually any subject population on any question.

A critical question is whether LLMs in fact faithfully replicate human personality and decision‐making. Several papers have conducted ``Turing tests,'' asking whether LLM behaviors and human behaviors can be  distinguished from one another. These papers find that LLMs broadly reproduce human tendencies with a few systematic differences.

For example, \citet{pmlr-v202-aher23a} ask LLMs to participate in the ultimatum game, in which one player proposes how to split a \$10 endowment, and the other decides whether to accept the proposal (where rejection leads to both players receiving nothing). The LLM distribution of offers and acceptance rates resembles human data, where in both cases offers of 50–100\% are  almost always accepted, while offers below 10\% are almost never accepted. 

\citet{Mei2024Turing} extend this analysis to a wider set of experimental games, including (among others) the dictator game, trust game, and finitely repeated Prisoner’s Dilemma. There is a large quantity of human data for these games,\footnote{\citet{Raman2025STEERME} evaluate LLMs on a battery of microeconomic questions, but their focus is on assessing its rationality rather than its similarity to human behavior.} and \citet{Mei2024Turing} use this to assess the likelihood of the LLM responses under the empirical human response distribution. They find that on average across games, a random LLM response is more probable under the empirical human distribution than a random human response, and in this sense the LLM distribution is human-like. 
(This finding should be interpreted with care: a deterministic distribution on the modal human action maximizes this ``human-like'' metric, but genuine mimicry may also require capturing heterogeneity across the distribution; we return to this point below.) Moreover, like humans, LLMs are sensitive to context and framing: \citet{Mei2024Turing} show that LLMs become more generous in the ultimatum and dictator games when told their choices will be observed by a third party, and \citet{Horton2023HomoSilicus} find that LLMs are more likely to choose an option when it is presented as the status quo. 

Nonetheless, some systematic differences remain. In a wisdom-of-the-crowds experiment, \citet{pmlr-v202-aher23a} instruct LLMs to predict the distribution of human responses to ten general-knowledge questions including ``How many bones does an adult human have?''. They find that the LLMs skew toward accurate answers, in one case predicting that a majority of simulated participants would answer all ten of the questions correctly. And across games with a distributional element (such as the dictator and ultimatum game), \citet{Mei2024Turing}  find that LLMs display greater cooperation and altruism than the median human. Since both tendencies likely reflect post-training objectives that encourage honesty and helpfulness (see, for example, \citet{bai2022constitutional}), these results suggest that LLMs would produce human-like behavior in the absence of interference, while also highlighting that post-training can systematically bias LLM outputs away from actual human distributions.

This line of research raises several fundamental questions.

\paragraph{What is the LLM's identity?}
A first challenge in using LLMs as experimental subjects is determining what identity---or identities---an LLM embodies. At least three possibilities emerge. First, we might view the LLM as a single, representative agent, with its answers corresponding to the modal response in the human population. This view is consistent with  \citet{Mei2024Turing}'s description of heterogeneity across LLM responses as ``within subject'' variation, and with their evaluation of human-likeness based on how far LLM responses differ from a typical human. If so, deploying LLMs as experimental subjects could reveal aggregate behavior and patterns, but would offer limited insight into heterogeneity across people.

Second, the LLM could be viewed as representing the full human population, with each response a random draw from the distribution of human behaviors and preferences. While LLMs do exhibit some randomness in their responses (which is tunable via the temperature parameter), \citet{Mei2024Turing} find that the variation across LLM responses falls far short of the heterogeneity observed among human subjects. The distributions also differ qualitatively: human response distributions have more distinct peaks, indicating multiple behavioral subgroups, whereas LLM response distributions are typically unimodal.\footnote{This is also consistent with the LLM having incorrect second-order beliefs about the human distribution, as demonstrated in \citet{pmlr-v202-aher23a}'s wisdom-of-the-crowds experiment.}

Third, we might view the LLM as a collection of identities, capable of mimicking a wide range of individuals when appropriately primed. \citet{pmlr-v202-aher23a} and \citet{Horton2023HomoSilicus} explicitly assume this perspective and induce variation in LLM responses by priming different identities---\citet{pmlr-v202-aher23a} by randomizing the names of the people whose responses the LLM is asked to simulate, and \citet{Horton2023HomoSilicus} by priming the LLM with different political identities. \citet{Mei2024Turing}, too, note the effects of identity priming: they find that in the ultimatum game, LLMs shift toward game‐theoretically  optimal strategies (accepting offers as low as \$1) when asked to act as a mathematician, and demand \$50 (out of \$100) in most cases when prompted as a legislator.

The variation in how LLMs are conceptualized across these papers highlights the need for a clearer account of LLM ``identity'' before they are used as experimental subjects at larger scale.  Moreover, researchers need ways to elicit and manage specific LLM identities, and ways to test hypotheses about those processes.

\paragraph{Can we trust LLM generalization?}

Suppose we establish in some principled way that LLM responses approximate human responses across a broad range of settings and data. A crucial question remains: will the LLM behave as humans do in new experiments, when confronted with problems and scenarios that are not represented in its training data?

This is critical to the value of LLMs for advancing behavioral science, as \citet{Horton2023HomoSilicus} envisions, since they are only useful if they can generate reliable responses for problems for which we don't already have data. In fact, \cite{Horton2023HomoSilicus} proposes that LLMs serve as sandbox laboratories in which one can test the consequences of counterfactual worlds and conditions  that would be difficult to credibly create with human subjects. He draws an analogy with economic theory, citing \citet{lucas1980methods}:

\begin{quote}
One of the functions of theoretical economics is to provide fully articulated, artificial economic systems that can serve as laboratories in which policies that would be prohibitively expensive to experiment with in actual economies can be tested at much lower cost.
\end{quote}

But there is a crucial difference: the predictions of a well‐specified theory are disciplined by the internal logic of the model. 
By contrast, LLMs offer no guarantee of such internal consistency. Their behavior in new tasks need not be (substantially) constrained by their behavior in prior tasks. In this respect, they resemble traditional machine learning models, and as with other ML systems, there is no reason to assume their predictions generalize.

It remains possible that an LLM has distilled some underlying representation of ``what it is to be human,'' and that this representation enforces cross‐context consistency. If so, the model’s behavior in new environments could indeed remain human‐like. But until we have evidence that its internal structure imposes such discipline, we should be cautious in interpreting out‐of‐sample LLM behaviors as indicative of what people will do.

\bigskip

Despite these notes of caution, the emerging evidence suggests that LLMs may fundamentally expand the scope and scale of experimental economics. Alongside this exploration, we need to develop rigorous frameworks for understanding their capabilities and limitations as models of humans. 

\subsection{LLMs for Creative and Heuristic Research Tasks} \label{sec:Heuristic}

A distinct possible use of LLMs is to automate parts of the research process that are informal and require creativity. This framing may be both surprising and provocative: creativity has long been viewed as a uniquely human capability, marking a clear divide between human and automated work. Moreover, because most existing LLM benchmarks emphasize tasks with objectively correct answers (e.g., solving International Math Olympiad problems), efforts to advance the technology have arguably focused more on improving reasoning than on supporting open‐ended idea generation. Yet LLMs are not formal symbolic machines; they are next‐word predictors trained to generate plausible continuations of text. This mode of generation often yields output that resembles creative, associative thinking. Whether such associative capabilities can make a substantive contribution to the heuristic and exploratory stages of research is an  empirical question.

\citet{Han2025MiningCausality} highlights one possible application: identifying instrumental variables (IVs). This process is inherently heuristic and relies largely on rhetoric---an IV’s validity is rarely established through formal proof but rather through persuasive argumentation about its relevance and exogeneity. LLMs can potentially automate this search. 

\citet{Han2025MiningCausality} demonstrates in classic economic problems that LLMs can not only reproduce known instrumental variables (IVs) but also propose new, plausible candidates. Consider the problem of estimating the causal effect of years of schooling on earnings. A valid IV in this context should be predictive of schooling attainment (relevance) but unrelated to any unobserved determinants of earnings (exogeneity), thus affecting earnings only through years of schooling.  When appropriately prompted, the LLMs recover standard instruments for this problem, including the distance from a student’s home to the nearest college and the number of siblings simultaneously attending college. The LLMs also suggest novel candidates, including campus crime rates and campus housing capacity. In a poll, experts found these novel IVs to be somewhat less valid than the established ones, but not by a wide margin.

In assessing findings of this kind, a natural concern is whether LLMs ``discover'' classic IVs through genuine reasoning or simply recall them from the training data. The novel candidates it produces offer some evidence of creative capability, but the true test going forward will be whether it can generate high‐quality IVs for genuinely new problems. 

Another creative endeavor in research is the generation of novel hypotheses. We already considered this goal in Section \ref{sec:ImproveModel}, where we observed that machine learning algorithms could be helpful in directing human attention, but ultimately required human input into the final formulation and naming of the hypothesis. LLMs offer an opportunity to automate the process more completely.

\citet{movva2024sparse} propose a method for generating natural-language hypotheses  using \emph{sparse autoencoders}. An autoencoder is a neural network that learns to compress an input vector $x$ into a lower‐dimensional hidden vector 
$h$ (the encoder) and then reconstruct it back into  $\hat{x}$ (the decoder). It is trained so that $\hat{x}$ is as close as possible to 
$x$. This forces the model to discover structure in the data---features that are sufficient to reconstruct the original input. A \emph{sparse} autoencoder adds a penalty so that only a small number of coordinates of $h$ are active for any given input. This discourages the network from simply memorizing inputs and instead encourages each hidden unit to capture a distinct, localized pattern in the data. The result is a set of learned features that tend to be more interpretable \citep{cunningham2023sparse}.

\citet{movva2024sparse}'s proposed approach is to: (1) train a sparse autoencoder on text embeddings to obtain a dictionary of interpretable features; (2) use regularized feature selection (see also Section \ref{sec:Interpretable}) to identify the features that are most predictive of the outcome; and (3) repeatedly prompt an LLM with prototypical high- and low-activation texts to produce concise natural-language labels for those features. On real datasets, the method recovers patterns such as: surprise-themed headlines receive more clicks, and speeches mentioning illegal immigration are more likely to be delivered by Republicans. 

While the specific hypotheses surfaced in these applications are more intuitive than surprising, the value of these approaches lies in making the creative step---i.e., mapping observed regularities into testable, natural-language hypotheses---more systematic and scalable. 
By contributing to these creative, heuristic stages of research, LLMs may complement more formal modeling and expand the toolkit available to experimental and theoretical economists.

\section{Machine Learning and Model-Building Beyond Economics} \label{sec:NotEcon}

This final section discusses related areas of work outside of economics: Section \ref{sec:Interpretable} considers the literature on interpretable machine learning, and Section \ref{sec:WorldModel} asks whether large language models have uncovered a model of the world.

\subsection{Interpretable ML} \label{sec:Interpretable}

The importance of interpretability in machine learning is powerfully illustrated by an incident described in \citet{Caruana2015Intelligible}. In a large-scale healthcare study, machine learning algorithms were trained to predict mortality risk for pneumonia patients. This prediction task had direct payoff consequences, since limited hospital capacity meant that risk assessment was critical for deciding which patients required admission and which should instead be treated as outpatients. Among the algorithms they considered, a neural network delivered the highest predictive accuracy. Yet it was ultimately rejected. Why?

One of the other algorithms the researchers trained was a more interpretable rule-based algorithm. When they examined this algorithm, they discovered it had learned to classify patients with a history of asthma as low-risk---a counterintuitive and potentially dangerous classification. But this pattern did in fact appear in the data: patients with asthma were typically admitted directly to intensive care and received aggressive treatment that succeeded in lowering their mortality rate. This finding raised the concern that the neural network might have learned other, equally problematic patterns that could not be as easily detected or corrected. To avoid this risk, the researchers chose to deploy a more transparent logistic regression model instead.

Episodes like this highlight the importance of understanding how a black-box model generates its predictions before entrusting it with important decisions. The field of interpretable machine learning (or ``explainable ML'') has developed a variety of techniques to “open up the black box” and examine its internal workings. These techniques can help determine whether a model has captured meaningful structure or is merely exploiting superficial correlations.

One widely used strategy is to assess which inputs have the greatest influence on a algorithm’s predictions. For example, the Shapley value of a feature  \citep{Shapley1951Value,Lundberg2017Unified} is its average impact on a algorithm's prediction. It is computed by comparing the algorithm's prediction with and without the feature across all choices for the remaining features, and averaging these changes in the algorithm's prediction. Intuitively, a feature is unimportant if the algorithm makes the same prediction with and without it, and the feature is important if the algorithm's predictions are responsive to the feature no matter what additional information it has. There are additionally notions of feature importance tailored to a specific model class. In LASSO regression, for instance, importance is often evaluated by the magnitude of the estimated coefficients, and for decision trees, importance of a given feature is measured by how much splitting on that feature reduces the outcome's entropy. 

Another major family of methods for explaining model predictions relies on perturbing a model’s inputs, internal representations, or training data \citep{Koh2017Understanding}, and observing how its predictions change. For example, in computer vision models, gradient-based saliency maps \citep{Simonyan2014DeepInsideConvNets} identify which pixels most influence a prediction by computing the sensitivity of the class score to changes in each pixel \citep{zhou2016learning,selvaraju2017gradcam}. More advanced approaches operate on deeper feature maps but project their results back into the original space of pixels for interpretability \citep{zeiler2014visualizing}. Such approaches can highlight semantically meaningful regions — for example, revealing that a sofa, table, and fireplace are key to classifying a scene as a ``living room '' \citep{zhou2016learning}.  

 Yet another class of approaches is to train a simpler, more interpretable model to approximate the behavior of a complex one. For example, \citet{Ribeiro2016WhyShouldITrustYou} propose finding explanations for the prediction $f(x)$ at a particular input $x$ by fitting an interpretable model in the neighborhood of $x$. Specifically, the analyst specifies a class of interpretable models (e.g., sparse linear models) and a measure of complexity for models in tht class (e.g., the number of nonzero coefficients). The analyst then searches within that class for the model best fits the black box locally around $x$, subject to a penalty on model complexity. The result is a low-complexity approximation of the black-box model's behavior around the point of interest. Similar ideas can be applied to learn a global surrogate model (such as a decision tree or rule list) that approximates the black box across the entire input space \citep{craven1996extracting,lakkaraju2016interpretable}.

\bigskip

This literature has informed several of the approaches discussed earlier, including \citet{hirasawa2022using}’s use of feature‐importance measures and \citet{LudwigMullainathan2024}’s use of gradient‐based image morphing, among others. But its primary motivation is different from that of economic theory. In particular, much of the work on interpretable ML is concerned with whether an algorithm’s predictions can be \emph{trusted}. In many applications, it is acceptable for the algorithm to base predictions on a complex function of the inputs that exploits statistical correlations, provided those correlations are not misleading (as in the pneumonia example). In such cases, the aim is to characterize how the model uses its inputs and how its internal representations relate to its predictions. By contrast, economic modeling seeks to uncover the mechanisms that generate outcomes, with the aim of guiding interventions at the level of policies, institutions, or social structures.\footnote{The emerging field of mechanistic interpretability shares this broader ambition of developing a holistic understanding, though the economic models considered here are by design much simpler than the neural networks this literature seeks to explain.} The next section asks whether algorithms may have attained such global understanding.

\subsection{``World Models''} \label{sec:WorldModel}

Economics is not the only discipline in which models and explanations of underlying mechanisms are valued alongside predictive accuracy. The question of whether large language models (LLMs) have learned such models---rather than serving solely as high-performing predictors---has recently become an active topic of research.

A central issue in this broader discussion is whether LLMs can uncover what some researchers call a \emph{world model}---a structured representation of the underlying rules or dynamics of a domain---or whether they are fundamentally limited to pattern-matching for prediction. 

There is no consensus gold standard  for how to evaluate whether the LLM possesses an understanding of a world model. One productive testbed has been to restrict attention to simple board games, where the ``world state'' is controllable and known. Specifically, suppose a large language model is given a large training set of play in the game, but is not given any information about the rules of the game.  Will the model eventually come to uncover these rules and to track the evolution of the board?

\citet{toshniwal2022chess} show that (when given sufficient data) a transformer is able to continue an initial sequence of chess moves with a valid following move. \citet{li2023emergent} replicate this finding for the game Othello, and moreover find evidence that the model has an internal representation of the state of the board.\footnote{This is achieved using nonlinear probes that seek to infer the board state from the internal network activations of the model.} These papers provide evidence that the model is not simply memorizing a large set of statistical associations, but has in fact learned something about the game.

\citet{vafa2024evaluating} argue that predicting legal next moves is not sufficient evidence that a model has learned the game. They create a dataset consisting of taxi rides in new York City and train a foundation model on these sequences. They find that the trained model has strikingly good route planning abilities: not only can it find valid routes between two points, but it usually identifies the shortest path. But when \citet{vafa2024evaluating} reconstruct the model's implicit street map of New York City, they find that it has little resemblance to the actual map, and moreover contains streets with impossible physical orientations.

This distinction between predictive capability and  an underlying world model is even starker in \citet{vafa2025what}. They show that a foundation model can predict planetary orbits with near-perfect accuracy, yet fail to uncover the Newtonian laws that govern them. When fine-tuned on a small dataset where success would require such a model to apply Newtonian mechanics, it instead extrapolates poorly—suggesting it has learned surface patterns tailored to the training data, rather than the deeper physical principles that would generalize.

Taken together, these papers highlight two points. First, different ways of evaluating whether an LLM has uncovered a ``world model'' can lead to different conclusions. Second, for some reasonable definitions of a world model,  high predictive performance does not, on its own, establish that a model has learned the correct underlying structure. Since the ``world model'' sought by these researchers closely parallels what this paper refers to as an ``economic model,'' this evidence outside of economics underscores the continuing importance of theory.

\section{Conclusion}

Artificial intelligence is likely to change the practice of scientific research in significant ways over the coming years. This paper suggests that machine learning methods can be complementary to traditional structural economic modeling.  Alongside the application of these methods, there is a need for clearer understanding of what they do, the conditions under which they are most effective, and how to interpret their outputs. This is particularly important for large language models (LLMs), which are likely to play an increasingly large role in the application of machine learning in economics.

\pagebreak

\end{document}